\documentclass[12 pt,aps,prd,showpacs]{revtex4-1}   
\usepackage{natbib}
\usepackage{amsmath}    
\usepackage{graphicx}   
\usepackage{epsfig}
\def\b{\begin{equation}}
\def\e{\end{equation}}

\def\kpara{{\bf k}_\parallel}
\begin{document}
\title{Inhibition of the dynamical Casimir effect with Robin boundary conditions}
\author{Andreson~L.~C.~Rego$^{1}$, B.~W.~Mintz$^{1,2}$, C.~Farina$^{1}$ and Danilo~T.~Alves$^{3}$}
\affiliation{
(1) - Instituto de F\'\i sica, Universidade Federal do Rio de Janeiro, Caixa Postal 68528, 21945-970, Rio de Janeiro, RJ, Brazil\\
(2) - Departamento de F\'\i sica Te\'orica, Universidade do Estado do Rio de Janeiro.
\\Rua S\~ao Francisco Xavier 524, Maracan\~a, 20550-900. Rio de Janeiro - RJ, Brazil.\\
(3)- Faculdade de F\'\i sica, Universidade Federal do Par\'a, 66075-110, Bel\'em, PA,  Brazil}
\date{February 22, 2013}

\begin{abstract}
We consider a real massless scalar field in $3+1$ dimensions satisfying a Robin boundary condition at a
nonrelativistic moving mirror. Considering vacuum as the initial field state, we compute explicitly the number of particles created per unit frequency and per unit solid angle, exhibiting in this way
the angular dependence of the spectral distribution. The well known cases of  Dirichlet and Neumann boundary conditions may be reobtained as particular cases from our results. We show that the particle creation rate can be considerably reduced (with respect to the Dirichlet and Neumann cases) for particular values of the Robin parameter. Our results extend for $3+1$ dimensions previous results found in the literature for $1+1$ dimensions. Further, we also show that this inhibition of the dynamical Casimir effect occurs for different angles of particle emission.
\end{abstract}

%

\maketitle
\section{Introduction}
\label{introduction}

The dynamical Casimir effect (DCE) basically consists of the emission of quanta from a moving body in vacuum due to its interaction with a quantized field
\cite{Moore-1970, Dewitt-PhysRep-1975, Fulling-Davies-PRS-1976-I, Davies-Fulling-PRS-1977-I, Davies-Fulling-PRS-1977-II}.
Another manifestation of the DCE, which is a direct consequence of the particle creation phenomenon if
we invoke the energy conservation law, is a radiation reaction force acting on the moving body. This dissipative force 
gives rise to an irreversible exchange of energy between the moving body and the quantized field. In other words, the energy
dissipated from the moving body is converted into real excitations of the quantized field, {\it i.e.}, real particles. One
can also understand the DCE in the opposite way, namely, from the fluctuation-dissipation theorem \cite{Nyquist-1928,Callen-Welton-1951}
and from the fact that the static Casimir force acting on a fixed body, though zero, has non-vanishing fluctuations \cite{Barton91}.
 In this scenario, one expects that a moving plate may be acted by a dissipative force (under certain circumstances) which is proportional to the
fluctuations of the Casimir force on the static plate \cite{Braginsky-Khalili-1991,Jaekel-Reynaud-1992,Barton94} (for the case of a
moving sphere see Ref. \cite{MaiaNeto-Reynaud-1993}).

However, the quantized field and the moving body can also exchange energy
reversibly, which means that the force exerted on the moving body acquires in this case a dispersive part, as it occurs when Robin
boundary conditions (BC) are considered \cite{Mintz-Farina-Maia-Neto-Robson-JPA-2006-I,Farina-BJP-2006}. During the first two decades
after the pioneering paper by Moore \cite{Moore-1970}, the calculations on the DCE were usually done with scalar fieds. The consideration
of electromagnetic fields was made by the first time in 1994 \cite{MaiaNeto-1994} (see also Ref(s) \cite{MaiaNeto-Machado-1996,MaiaNeto-Mundurain-1998}), and
since then great attention has been devoted to the DCE. Detailed reviews on the DCE can be found in Refs.
\cite{Dodonov-Revisao,PAMN-EtAl-Revisao}. Several experimental proposals to observe the DCE have been made in the last years. We shall briefly comment on a couple of them (for more details see, for instance, Dodonov's paper \cite{V-V-Dodonov-Phys-Scrip-2010}).

The so called motion induced radiation (MIR) experiment \cite{Braggio-Agnesi} is based on the simulation of a mirror's motion by changing the reflectivity of a semi-conductor by irradiating it with appropriate laser pulses, an ingenious idea firstly introduced by Yablonovitch in 1989 \cite{Yablonovitch-1989} in a paper where the main concern was to propose ways of simulating highly accelerated frames in order to enhance the Unruh radiation.  A few years later, this same idea of creating a dense electron-hole plasma in a thin semiconductor by irradiating it with laser pulses, was also discussed by Lozovik {\it et al} \cite{LozovikEtAl-1995}. Though there are many promissing aspects in the MIR experiment, it is worth mentioning that the MIR experimentalists may have to deal with some difficulties. A first one is related to the limitations in the signal-to-noise ratio present in their experiment caused by thermal effects, in case they run the experiment at 4.6 K, as pointed out by  Kim {\it et al} \cite{Kim-Brownell-Onofrio-EPL-2007}.
A second one is the influence of damping in a parametric amplification process. And in the MIR experiment, the electric permittivity of the semiconductor slab after excitation by the laser pulse acquires a non-negligible imaginary part
\cite{Dodonov-2005}, so that dissipation effects are inevitable. In this case, as shown by Dodonov, damping plays an important role and the emergence of a superchaotic quantum state may occur leading to a highly superPoissonian statistics for the distribution function of quanta \cite{Dodonov-2009}.

Another interesting proposal was made by  Kim {\it et al} \cite{Kim-Brownell-Onofrio-PRL-2006}.
They suggest that an indirect measurement of the dynamical Casimir photons generated by means of the mechanical motion of a film bulk acoustic resonator is detected with the aid of a  superadiance mechanism.
More recently, Dezael and Lambrecht \cite{Dezael-Lambrecht-EPL-2010} proposed that a (dynamical) Casimir-like radiation may arise from an effective motion of mirrors obtained by the interactions of an optical parametric oscillator with a thin non-linear crystal slab inside.
In 2011, Kawakubo and Yamamoto \cite{Kawakubo-Yamamoto-PRA-2011} proposed that photons could be generated by means of a non-stationary plasma mirror, being the photons detectable by an excitation process of Rydberg atoms through the atom-field interaction.
Still in 2011, Faccio and Carusotto \cite{Faccio-Carusotto-EPL-2011} proposed a photon generation mechanism in the near-infrared domain obtained by a train of laser pulses applied perpendicularly to a cavity, made of non-linear optical fiber, which modulates in time the refractive index of the medium filling the cavity.

Finally, forty years after its theoretical prediction made by Moore \cite{Moore-1970}, the first experimental
observation of the DCE was announced by Wilson and collaborators \cite{C-Wilson-et-al-Nature-2011}, in an experiment where these authors use a superconducting circuit consisting of a unidimensional coplanar transmission line with a tunable electrical length. The change of the electrical length is performed by modulating the inductance of a superconducting quantum interference device fixed at one end of the transmission line. This modulation is
achieved with the aid of a time-dependent magnetic flux through the superconducting quantum interference device. The electromagnetic field along the transmission line is described in terms of a field operator given by a scalar field $\phi(t,x)$ obeying a massless Klein-Gordon equation in 1+1 dimensions and submitted to a Robin BC with a time-dependent Robin parameter
$\gamma(t)$ in the following manner \cite{J-R-Johansson-G-Johansson-C-Wilson-F-Nori-PRL-2009}:
\begin{equation}
\phi(t,0) \approx \gamma(t){(\partial_x\phi)}(t,0).
\label{eq:Robin_BC_static}
\end{equation}
This kind of BC was also considered in Ref(s) \cite{Silva-Farina-PRD-2011,Farina-Silva-Rego-Alves-IJMPCS-2012}.
In the context of the DCE, Robin BC appeared for the first time in the
papers by Mintz and collaborators \cite{Mintz-Farina-Maia-Neto-Robson-JPA-2006-I,Mintz-Farina-Maia-Neto-Robson-JPA-2006-II},
who investigated a real massless scalar field $\phi(t,x)$ satisfying a Robin BC at a moving plate when observed from an inertial frame in which the plate is instantaneously at rest (we shall refer to this frame as tangential frame), according to the formula
\begin{equation}
\phi^{\prime}(t^{\prime},x^{\prime}) =\gamma{(\partial_{x^{\prime}}\phi^{\prime})}(t^{\prime},x^{\prime}),
\label{eq:Robin_BC_dynamical}
\end{equation}
where the prime superscript is to remind us that the BC is taken
in a tangential frame and $\gamma$ is now a time-independent
Robin parameter.

For an oscillating mirror that imposes Dirichlet BC on a scalar field in $1+1$ dimensions, the total particle creation rate is a
monotonic function of the mechanical frequency of the mirror \cite{Lambrecht-PRL-1996}. However, it was shown in
\cite{Mintz-Farina-Maia-Neto-Robson-JPA-2006-II} that this is not always the case when the field obeys a Robin BC given by Eq. (\ref{eq:Robin_BC_dynamical}). For a given value of $\gamma$, there is an interval in which an increase in the oscillation frequency of the mirror leads to a decrease in the particle emission. This can be understood as a kind of ``decoupling'' between the mirror and some of the field modes.

One important question that remains is whether or not this interesting effect still occurs  in $3+1$ dimensions.
This is not a trivial issue, once the phase space available for the field in $d=3+1$ is much larger than that in $1+1$ dimensions. To answer this question is the main purpose of our paper and we shall do that by considering a real massless scalar field in $3+1$ dimensions satisfying a Robin BC at  a non-relativistic moving mirror (in the tangential frame). As far as we know, all papers about DCE with Robin BC deal with  models in 1+1 dimensions, except one \cite{Diogo-Farina-Procceeding-2011}, which did not discuss the problem we are interested here.

An extra motivation to study this model is the connection between
the DCE for a real massless scalar field in $3+1$ dimensions
and the DCE for the electromagnetic field interacting with a perfectly conducting plate.
The latter problem can be separated  into two problems: a vector
potential representing the transverse electric (TE) polarization, which  is associated to
a Dirichlet BC, and a vector potential representing the transverse magnetic (TM) polarization, which is associated to a Neumann BC. Since the parameter $\gamma$ allows a continuous interpolation
between Dirichlet and Neumann BC, we expect that in the limit $\gamma\rightarrow0$ the DCE for the massless scalar field coincides with the TE polarization contributionto the electromagnetic DCE, while for $\gamma\rightarrow\infty$ the TM polarization contribution to the electromagnetic DCE is recovered.

The structure of this paper is as follows. In Sec. II, we calculate the relations between the creation and annihilation field operators in the remote past (``in'' operators) and in the far future (``out'' operators), in the Heisenberg picture. In Sec. III, we show our results for the particle emission rate for a specific but typical motion of the mirror. Finally, in Sec. IV, we discuss our results and make a few comments on their possible consequences.

%
\section{Bogoliubov Transformation and Particle Spectrum}
\label{transf-bogoliubov}
%

Let us consider - in the Heisenberg picture - a massless scalar field $\phi$ in $3+1$ dimensions
written in terms of the time-dependent operators $a^\dagger(t,\bf k)$ and $a(t,\bf k)$,
where ${\bf k}$ is the wave vector.
In the distant past, these operators are relabeled as the creation and annihilation operators $a_{in}^\dagger({\bf k})$ and $a_{in}({\bf k})$,
whereas in the far future they are relabeled as $a_{out}^\dagger({\bf k})$ and $a_{out}({\bf k})$.
The time evolution of the operators $a^\dagger$ and $a$ depends on the interaction between the
field with an external agent (modeled, in the present paper, by a moving boundary). The
``out'' operators can be expressed as a combination of the ``in'' operators
via Bogoliubov transformations \cite{Bogoliubov}
\begin{equation}\label{eq:bogoliubov_transformations}
 a_{out}({\bf k}) = \alpha({\bf k}) a_{in}({\bf k}) + \beta({\bf k}) a_{in}^\dagger({\bf k}),
\end{equation}
where $\alpha$ and $\beta$ are named Bogoliubov coefficients. 
Assuming that in the remote past the system is in the vacuum state $|0\rangle$, the spectral density of the created particles after the movement of the mirror has ceased is given by
\begin{equation}\label{eq:spectral-density}
 \frac{dN}{d^3k}({\bf k}) = \frac{1}{\left(2\pi\right)^3}\langle0| a_{out}^\dagger({\bf k}) a_{out}({\bf k}) |0\rangle
 =|\beta({\bf k})|^2.
\end{equation}
Notice that there will be particle production if,
and only if, $\beta({\bf k})\not=0$, i.e., if the annihilation operator $a_{out}({\bf k})$ is ``contaminated''
by the creation operator $a_{in}^\dagger({\bf k})$.
The relation between the ``in'' and ``out'' operators can be calculated exactly for the case  $1+1$ dimensions
\cite{Davies-Fulling-PRS-1977-II} with Dirichlet or Neumann BC. However, only approximate approaches are
currently known for higher dimensional space-times. If the movement of the mirror is non-relativistic and
has a small amplitude, the perturbative method introduced by Ford and Vilenkin \cite{Ford-Vilenkin-PRD-1982} applies.
In the following, we will use the Ford-Vilenkin approach to find the Bogoliubov transformation
for the massless scalar field satisfying the wave equation
\begin{equation}\label{eq:dAlembert}
 \frac{\partial^2\phi}{\partial t^2} - \nabla^2\phi = 0
\end{equation}
and obeying a Robin BC imposed by a moving mirror.
We start writing the field operator as the field under static BC plus a
perturbation, that is
\begin{equation}\label{eq:Ford-Vilenkin_ansatz}
 \phi(t,{\bf r}) = \phi_0(t,{\bf r}) + \delta\phi(t,{\bf r}),
\end{equation}
where we assume that the undisturbed field $\phi_0$ obeys the static boundary condition BC
\begin{equation}\label{eq:static_BC}
\Bigl[\phi_0\left(t,{\bf r}\right)-\gamma\partial_{z}\phi_0\left(t,{\bf r}\right)\Bigr]_{z=0}=0,
\end{equation}
where $\gamma$ is a time-independent parameter.

The perturbation $\delta\phi$ (which is assumed to be small) gives the first order contribution to the total  field $\phi(t,{\bf r})$ caused by the movement of the mirror and therefore will be responsible for the emergence of the DCE.

Let $z=\delta q(t)$ be the position of the mirror at a given instant. We assume that, in the
lab reference frame, the mirror starts at rest at $z=0$, undergoes a given prescribed movement and finally
settles down at $z=0$ for large times. The perturbation $\delta\phi$ will be small as long as the speed of
the mirror is continuous with continuous derivatives and is much smaller than the speed of light,
$|\delta \dot q(t)|\ll1$, in natural units. Once we consider a bounded movement for the mirror, we also
require that it possesses a small amplitude. More specifically, this means that $\vert\delta q\vert\ll 1/\omega_0$, where $\omega_0$ is the dominant mechanical frequency and this assumption will allow us to neglect terms
${\cal O}[(\delta q)^2]$, ${\cal O}[(\delta \phi)^2]$, ${\cal O}[\delta\phi\delta q]$ and
${\cal O}[(\delta \dot q)^2]$.

We shall follow the same procedure as in \cite{Mintz-Farina-Maia-Neto-Robson-JPA-2006-II}.
At the tangential frame  at a given instant, the BC is given by
\begin{equation}\label{eq:Robin_BC_comoving_frame}
 \phi'(t',{\bf r}') = \gamma' \frac{\partial\phi'}{\partial z'}(t',{\bf r}'),
\end{equation}
where $\gamma'$ is a $t'$-independent parameter. This BC can be cast in the lab frame with the
help of the appropriate Lorentz transformations. For non-relativistic velocities,
one may expand in first order in $\delta\dot q(t)$ to find
\begin{equation}\label{eq:first_order_Robin-BC}
\Bigl[\partial_{z}\phi\left(t,{\bf r}\right)+\delta\dot{q}(t)\partial_{t}\phi\left(t,{\bf r}\right)-\gamma^{-1}\phi\left(t,{\bf r}\right)\Bigr]_{z=\delta q(t)}=0,
\end{equation}
where, in this approximation, we consider that the Robin parameter is not affected
by the Lorentz transformation, so that $\gamma'=\gamma$.

We substitute (\ref{eq:Ford-Vilenkin_ansatz}) in (\ref{eq:first_order_Robin-BC}) and expand the result thus obtained
in the small parameters retaining only terms up to linear order
in $\delta q$ and $\delta\phi$, finding
%
\begin{eqnarray}\label{eq:inhomog_bc}
\partial_{z}\delta\phi\left(t,{\bf r}\right)|_{z=0} \;-\;    
 \gamma^{-1}\delta\phi\left(t,{\bf r}\right)|_{z=0} \;\;&=&\;\;
\delta q(t)\gamma^{-1}\partial_{z}\phi_0(t,{\bf r})|_{z=0} 
\;-\; \delta q(t)\partial_{z}^2\phi_0(t,{\bf r})|_{z=0} 
\cr &&
\;-\; \delta\dot{q}(t)\partial_{t}\phi_0\left(t,{\bf r}\right)|_{z=0}.
\end{eqnarray}
%
Therefore, the perturbation $\delta\phi(t,{\bf r})$ obeys a time-dependent BC at a fixed position $(z=0)$, which is associated to the motion
of the mirror via $\delta q(t)$.

It is now convenient that we express the field in the frequency domain using a Fourier transform, as follows.
Let us define $\Phi(\omega,{\bf k}_{\parallel};z)$ as the Fourier transform of the field $\phi(t,{\bf r})$,
so that
\begin{equation}\label{eq:field_FourierTransf}
\Phi(\omega,{\bf k}_{\parallel};z):=\int_{-\infty}^{\infty} dt \int d^2{\bf r}_{\parallel}
     e^{i \omega t} e^{- i {\bf k}_{\parallel}\cdot{\bf r}_{\parallel}} \phi(t,{\bf r}).
\end{equation}
Notice that $\Phi(\omega,{\bf k}_{\parallel};z)$ obeys the Helmholtz equation
\begin{equation}
\left(\omega^2-k^2_{\parallel}+\partial_{z}^2\right)\Phi(\omega,{\bf k}_{\parallel};z)=0.
\label{helholtz-total}
\end{equation}

Once the massless free scalar field is a solution of the d'Alembert equation (\ref{eq:dAlembert}),
\begin{equation}
k_z=\left[(\omega+i\epsilon)^2-k_{\parallel}^2\right]^{1/2}, \;\;
\text{with} \;\; \epsilon \rightarrow 0^{+}.
\label{kz}
\end{equation}
With this definition, $k_z$ is a complex function of $\omega$, with a branch cut along the real axis between
$-k_{\parallel}<\omega<k_{\parallel}$ \cite{MaiaNeto-Machado-1996}, where $k_\parallel = \vert{\bf k}_\parallel\vert$.

The undisturbed field is the solution of the wave equation (\ref{eq:dAlembert}) subject to the static
BC (\ref{eq:static_BC}):
\begin{eqnarray}
\phi_0(t,{\bf r}) &=& \int_{-\infty}^{\infty} d^2{\bf k}_{\parallel} \int_{0}^{\infty} dk_z
\sqrt{\frac{1}{4\pi^3\left|\omega_{{\bf k}}\right|}\left(\frac{1}{1+\gamma^2 k_z^2}\right)}
\Biggl[a({\bf k})e^{- i \omega_{{\bf k}} t} e^{i {\bf k}_{\parallel}\cdot{\bf r}_{\parallel}}
+a^\dag({\bf k})e^{ i \omega_{{\bf k}} t} e^{-i {\bf k}_{\parallel}\cdot{\bf r}_{\parallel}}\Biggr]
\times \cr && 
\Biggl[\sin(k_{z}z)+\gamma k_z \cos(k_{z}z)\Biggr],
\label{eq:undisturbed_field_solution}
\end{eqnarray}
where the field normalization is chosen so that the creation and annihilation operators obey the commutation
relations
\begin{equation}
 \left[ a({\bf k}),a^\dag(\bf k)'\right] = (2\pi)^3\delta({\bf k}-{\bf k}').
\end{equation}

We can now use (\ref{eq:field_FourierTransf}) to write down the Fourier transform of the undisturbed field,
\begin{eqnarray}
\Phi_0(\omega,{\bf k}_{\parallel};z) &=& A(\omega,k_z;\gamma) \Bigl[\sin(k_{z}z)+\gamma k_z \cos(k_{z}z)\Bigr]
\!\!\Bigl[a({\bf k})\Theta(\omega)-a^\dag(-{\bf k})\Theta(-\omega)\Bigr]\Theta(k_z^2),\;\,
\label{campo-nao-perturbado-robin-recip}
\end{eqnarray}
with the normalization factor
\begin{equation}
A(\omega,k_z;\gamma)=\sqrt{\frac{16\pi^3\left|\omega\right|}{k_z^2}\left(\frac{1}{1+\gamma^2 k_z^2}\right)}.
\label{constante-robin-recip}
\end{equation}

The Fourier transformations of the perturbation  $\delta\phi(t,{\bf r})$ and the mirror's law of motion $\delta q(t)$ are written as
\begin{equation}
\delta\Phi(\omega,{\bf k}_{\parallel};z) =
{\int_{-\infty}^{\infty} dt \int d^2{\bf r}_{\parallel} e^{i \omega t} e^{- i {\bf k}_{\parallel}.{\bf r}_{\parallel}} \delta\phi(t,{\bf r})}
\end{equation}
and
\begin{equation}
\delta Q(\omega) =
{\int_{-\infty}^{\infty} dt\, e^{i \omega t} \delta q(t)}.
\label{eq:mirror_law_motion}
\end{equation}

Notice that $\delta\Phi(\omega,{\bf k}_{\parallel};z)$ obeys the Helmholtz equation
\begin{equation}
\left(\omega^2-k^2_{\parallel}+\partial_{z}^2\right)\delta\Phi(\omega,{\bf k}_{\parallel};z)=0.
\label{helholtz-perturbacao}
\end{equation}
Recalling that we must take only the solution that propagates outwards from the moving mirror, the solution of the
previous equation for $z>0$ is given by
\begin{equation}\label{eq:pert_solution_Helmholtz}
\delta\Phi(\omega,{\bf k}_{\parallel};z)=\frac{1}{ik_z}\partial_{z}\delta\Phi(\omega,{\bf k}_{\parallel};0) e^{ik_{z}z}.
\end{equation}

We now notice also that $\delta\Phi(\omega,{\bf k}_{\parallel};z)$ obeys the BC given by the Fourier
transform of (\ref{eq:inhomog_bc}), namely,
%
\begin{eqnarray}\label{eq:inhomog_bc_fourier}
\partial_z\delta\Phi(\omega,{\bf k}_{\parallel};0) -  
\gamma^{-1}\delta\Phi(\omega,{\bf k}_{\parallel};0) &=& 
\gamma^{-1}\int_{-\infty}^{\infty} \frac{d\omega^{\prime}}{2\pi}\delta Q(\omega-\omega^{\prime})
\Bigl[\partial_z \Phi_0(\omega^{\,\prime},{\bf k}_{\parallel};0) + \Bigr.
\cr &&
\Bigl.
\Bigl(\gamma \omega \omega^{\prime} - \gamma\kpara^2\Bigr)\Phi_0(\omega^{\,\prime},{\bf k}_{\parallel};0)\Bigr]
\end{eqnarray}
%

Once the Fourier transform of the Ford-Vilenkin ansatz (\ref{eq:Ford-Vilenkin_ansatz}) is
\begin{equation}
\Phi(\omega,{\bf k}_{\parallel};z)=\Phi_{0}(\omega,{\bf k}_{\parallel};z)+\delta\Phi(\omega,{\bf k}_{\parallel};z),
\label{fv-recip}
\end{equation}
we now need to find an expression for $\Phi(\omega,{\bf k}_{\parallel};z)$ that allows us to relate the Fourier transforms of the field in the remote past and distant future. This can be done straightforwardly by using Green's functions, as in \cite{Mintz-Farina-Maia-Neto-Robson-JPA-2006-II}.

Let us start with the one-dimensional version of Green's identity,
$\partial_z\{g[\partial_{z}f]-f[\partial_{z}g]\}=g[\partial_{z}^{2}f]-f[\partial_{z}^{2}g]$.
We now identify the function $f$ in this formula as the perturbation $\delta\Phi$
and the function $g$ as the Green's function of the Helmholtz operator, i.e., a function
such that $\left(\omega^2-k^2_{\parallel}+\partial_{z}^2\right)g(\omega,{\bf k}_{\parallel};z,z^{\prime})=\delta(z-z^{\prime})$.
After integration by parts, we can show that
\begin{eqnarray}
\delta\Phi(\omega,{\bf k}_{\parallel};z^{\prime}) &=&
\Bigl[ g(\omega,{\bf k}_{\parallel};0,z^{\prime})\partial_{z}\delta\Phi(\omega,{\bf k}_{\parallel};0) 
\partial_{z}g(\omega,{\bf k}_{\parallel};0,z^{\prime})\delta\Phi(\omega,{\bf k}_{\parallel};0)\Bigr].
\label{campo-perturbado-recip-novo}
\end{eqnarray}

We may freely define the Green's function $g$ so that it obeys the Robin BC at the surface of the static mirror, i.e., 
$g(\omega,{\bf k}_{\parallel};0,z^{\prime}) = \gamma\partial_{z}g(\omega,{\bf k}_{\parallel};0,z^{\prime})$.
Then, the identity (\ref{campo-perturbado-recip-novo}) leads to
\begin{eqnarray}
&\delta&\!\Phi(\omega,{\bf k}_{\parallel};z^{\prime})
g(\omega,{\bf k}_{\parallel};0,z^{\prime})
\!\Bigl[\partial_{z}\delta\Phi(\omega,{\bf k}_{\parallel};0) - 
\gamma^{-1}\delta\Phi(\omega,{\bf k}_{\parallel};0)\!\Bigr]\;
\label{campo-perturbado-em-termos-das-cc-recip}
\end{eqnarray}
and, using Eq. (\ref{eq:inhomog_bc_fourier}), we can write the perturbation as
%
\begin{eqnarray} 
\delta\Phi(\omega,{\bf k}_{\parallel};z^{\prime}) &=&
\gamma^{-1}g(\omega,{\bf k}_{\parallel};0,z^{\prime}) \int_{-\infty}^{\infty} \frac{d\omega^{\prime}}{2\pi}\delta Q(\omega-\omega^{\prime})
\Bigl[\partial_z \Phi_0(\omega^{\,\prime},{\bf k}_{\parallel};0) + \Bigr.
\cr && \Bigl. 
\Bigl(\gamma \omega \omega^{\prime} - \gamma\kpara^2\Bigr)\Phi_0(\omega^{\,\prime},{\bf k}_{\parallel};0)\Bigr]
\label{campo-perturbado-recip-em-termos-do-campo-nao-perturbado}
\end{eqnarray}
%

Replacing (\ref{campo-perturbado-recip-em-termos-do-campo-nao-perturbado}) in (\ref{fv-recip}) one finds
%
\begin{eqnarray}\label{campo-total-campo-np}
\Phi(\omega,{\bf k}_{\parallel}; z^{\prime}) &=&
\Phi_{0}(\omega,{\bf k}_{\parallel};z^{\prime}) + \frac{1}{\gamma}g(\omega,{\bf k}_{\parallel};0,z^{\prime})\int_{-\infty}^{\infty} \frac{d\omega^{\prime}}{2\pi}
\delta Q(\omega-\omega^{\prime})
\Bigl[\partial_z \Phi_0(\omega^{\,\prime},{\bf k}_{\parallel};0) + \Bigr.
\cr && \Bigl.
\Bigl(\gamma \omega \omega^{\prime} - \gamma\kpara^2\Bigr)\Phi_0(\omega^{\,\prime},{\bf k}_{\parallel};0)\Bigr].
\end{eqnarray}
%

The field $\Phi(\omega,{\bf k}_{\parallel};z^{\prime})$ can be written in two different forms.
The first one involves $\Phi_{in}$, the Fourier transform of the unperturbed field
$\phi_{0}(t,{\bf r}_{\parallel};z^{\prime})$ in the distant past, that is, the ``in'' field,
to which we associate retarded Green's functions. Then,
%
\begin{eqnarray}
\Phi(\omega,{\bf k}_{\parallel}; z^{\prime}) &=&
\Phi_\text{in}(\omega,{\bf k}_{\parallel};z^{\prime}) + \frac{1}{\gamma}g_\text{ret}(\omega,{\bf k}_{\parallel};0,z^{\prime})\!\int_{-\infty}^{\infty} \frac{d\omega^{\prime}}{2\pi}
\delta Q(\omega-\omega^{\prime}) 
\Bigl[\partial_z \Phi_0(\omega^{\,\prime},{\bf k}_{\parallel};0) + \Bigr.
\cr && \Bigl.
\Bigl(\gamma \omega \omega^{\prime} - \gamma\kpara^2\Bigr)\Phi_0(\omega^{\,\prime},{\bf k}_{\parallel};0)\Bigr],
\label{campo-total-campo-np-in}
\end{eqnarray}
%
where $\Phi_\text{in}(\omega,{\bf k}_{\parallel};z^{\prime})$ is the Fourier transform of $\phi_\text{in}(t,{\bf r})$, which is simply the unperturbed field at the remote past,
\begin{equation}
\phi_{0}(t,{\bf r}) \mathop{\sim}_{t\rightarrow-\infty} \phi_\text{in}(t,{\bf r}).
\label{campo-passado-remoto}
\end{equation}
Analogously, one can use advanced Green's functions to express the field $\Phi_{0}(\omega,{\bf k}_{\parallel};z^{\prime})$ using the ``out'' field configuration, as follows,
%
\begin{eqnarray}\label{campo-total-campo-np-out}
\Phi(\omega,{\bf k}_{\parallel}; z^{\prime}) &=&
\Phi_\text{out}(\omega,{\bf k}_{\parallel};z^{\prime}) + \frac{1}{\gamma}g_\text{adv}(\omega,{\bf k}_{\parallel};0,z^{\prime})
\int_{-\infty}^{\infty}\frac{d\omega^{\prime}}{2\pi}
\delta Q(\omega-\omega^{\prime}) 
\Bigl[\partial_z \Phi_0(\omega^{\,\prime},{\bf k}_{\parallel};0) + \Bigr.
\cr && \Bigl.
\Bigl(\gamma \omega \omega^{\prime} - \gamma\kpara^2\Bigr)\Phi_0(\omega^{\,\prime},{\bf k}_{\parallel};0)\Bigr],
\end{eqnarray}
%
where  $\Phi_\text{out}(\omega,{\bf k}_{\parallel};z^{\prime})$ is the Fourier transform of
\begin{equation}
\phi_{0}(t,{\bf r}) \mathop{\sim}_{t\rightarrow\infty} \phi_\text{out}(t,{\bf r}).
\label{campo-futuro-distante}
\end{equation}

From Eqs. (\ref{campo-total-campo-np-out}) and (\ref{campo-total-campo-np-in}), we can relate the
``in'' and ``out'' field operators through the advanced and retarded Green's functions,
%
\begin{eqnarray}\label{campo-in-out}
\Phi_\text{out}(\!\omega\!, {\bf k}_{\parallel}; z^{\prime}) \;&=&\; 
\Phi_\text{in}(\omega,{\bf k}_{\parallel};z^{\prime}) \;+  
\frac{1}{\gamma}\Bigl[g_\text{ret}(\omega,{\bf k}_{\parallel};0,z^{\prime})-g_\text{adv}
(\omega,{\bf k}_{\parallel};0,z^{\prime})\Bigr]
\times \cr &&
\!\int_{-\infty}^{\infty}\!\!\! \frac{d\omega^{\prime}}{2\pi}
\delta Q(\omega-\omega^{\prime})
\!\Bigl[\partial_z \Phi_0(\omega^{\,\prime},{\bf k}_{\parallel};0) + 
\!\Bigl(\gamma \omega \omega^{\prime} - \gamma\kpara^2\Bigr)\Phi_0(\omega^{\,\prime},{\bf k}_{\parallel};0)\!\Bigr]\;
\end{eqnarray}
%

Using the solution (\ref{campo-nao-perturbado-robin-recip}) of the Helmholtz equation for the field subject to Robin
BC, one can reexpress (\ref{campo-in-out}) as
\begin{eqnarray}
\Phi_\text{out}(\omega,{\bf k}_{\parallel};z^{\prime}) &=&
\Phi_\text{in}(\omega,{\bf k}_{\parallel};z^{\prime}) + \left(\frac{2i}{1+\gamma^{2}k_{z}^{2}}\right)
\Bigl[\sin(k_{z}z^{\prime})+\gamma k_z \cos(k_{z}z^{\prime})\Bigr]
\times \cr &&
\int_{-\infty}^{\infty} \frac{d\omega^{\prime}}{2\pi} 
A(\omega^{\prime},k_{z}^{\prime};\gamma)
\delta Q(\omega-\omega^{\prime})k_{z}^{\prime}
\Bigl(1-\gamma^2\kpara^2+\gamma^2\omega\omega^{\prime}\Bigr)
\times \cr &&
\Bigl[a({\bf q})\Theta(k_{z}^\prime)-a^\dag(-{\bf q}) 
\Theta(-k_{z}^\prime)\Bigr]\Theta({k_{z}^{\prime2}}),
\label{campo-in-out-novo}
\end{eqnarray}
where we defined
\begin{equation}
 {\bf q}:=\kpara + k_z^\prime\hat z,
\end{equation}
with
\begin{equation}
 k_z^\prime=\left[(\omega^\prime+i\epsilon)^2 - \kpara^2\right]^{1/2}.
\end{equation}
Once again, using (\ref{campo-nao-perturbado-robin-recip}) it is straighforward to show that
\begin{eqnarray}\label{eq:Bogoliubov_transf}
a_\text{out}\!(\!{\bf k}) &=& a_\text{in}({\bf k}) + \frac{2ik_z}{\sqrt{\left|\omega\right|(1+\gamma^{2}k_{z}^{2})}}
\int_{-\infty}^{\infty} \frac{d\omega^{\prime\prime}}{2\pi}
{\sqrt\frac{\left|\omega^{\prime\prime}\right|}{(1+\gamma^{2}k_{z}^{\prime\prime2})}}
\delta Q(\omega-\omega^{\prime\prime})
\Bigl(1-\gamma^2\kpara^2+\gamma^2\omega\omega^{\prime\prime}\Bigr)
\times \cr &&
\Bigl[a_\text{in}({\bf q}^{\prime\prime})\Theta(k_{z}^{\prime\prime})-a_\text{in}^\dag(-{\bf q}^{\prime\prime})\Theta(-k_{z}^{\prime\prime})\Bigr]\Theta({k_{z}^{\prime\prime2}}),
\label{operadores-ca-in-out}
\end{eqnarray}
with ${\bf q}^{\prime\prime}=\kpara+k_z^{\prime\prime}\hat z$. The last expression is a linear relation
between the annihilation field operator in the far future with the creation and annihilation operators
at the remote past. That is, (\ref{operadores-ca-in-out}) is the Bogoliubov transformation we were looking
for.

As advertised, the vacuum state is annihilated by the ``in'' operator $a_{in}({\bf k})$, but not by the
``out'' operator $a_{out}({\bf k})$ due to the presence of $a_{in}^\dag({\bf k})$ in the r.h.s. of
(\ref{operadores-ca-in-out}).  This clearly indicates a nonzero particle creation in $3+1$ dimensions 
due to the DCE for a massless field subject to Robin BC at a moving mirror.

After some straightforward manipulations, from Eq. (\ref{eq:spectral-density}) and using the Bogoliubov transformation (\ref{eq:Bogoliubov_transf}), we can finally calculate the spectral density of created particles:
\begin{eqnarray}
\frac{dN({\bf k})}{d^3{\bf k}} &=&
\frac{A}{\left(2\pi\right)^3}\frac{4k_z^2}{|\omega|(1+\gamma_0k_z^2)}
\int_{|\kpara|}^\infty\frac{d\omega^\prime}{2\pi}k_z^\prime
\frac{|\delta Q(\omega+\omega^\prime)|^2}{1+\gamma_0k_z^{\prime2}}
\Bigl(1-\gamma^2\kpara^2 - \gamma^2\omega\omega^\prime\Bigr)^2,
\label{eq:angular_frequency_spectrum}
\end{eqnarray}
%
where $A$ is the area of the moving plate.
Taking into account the axial symmetry of the problem, it is interesting to rewrite
(\ref{eq:angular_frequency_spectrum}) in terms of the variables
\begin{equation}
 k_z = \omega\cos\theta\;\;\;\mbox{and}\;\;\;|\kpara| = \omega\sin\theta,
\end{equation}
with $\omega=|{\bf k}|$ and $\theta\in[0,\pi/2]$. Analogously, we have
\begin{equation}
 k_z^\prime = \sqrt{\omega^{\prime2} - \kpara^2} = \sqrt{\omega^{\prime2} - \omega^2\sin^2\theta},
\end{equation}
which leads us to our expression for the particle spectrum per unit area,
per unit solid angle
\begin{eqnarray}
 \frac{1}{A}\frac{dN}{d\omega d\Omega} &=&
\frac{1}{\left(2\pi\right)^3}
\frac{4\omega^3\cos^2\theta}{1+\gamma^2\omega^2\cos^2\theta}
\int_{\omega\sin\theta}^\infty\frac{d\omega^\prime}{2\pi}
\sqrt{\omega^{\prime2} - \omega^2\sin^2\theta} \;
\frac{|\delta Q(\omega+\omega^\prime)|^2}{1+\gamma^2(\omega^{\prime2}-\omega^2\sin^2\theta)}
\times \cr &&
\Bigl(1-\gamma^2\omega^2\sin^2\theta - \gamma^2\omega\omega^\prime\Bigr)^2.
\label{eq:spectrum_general_movement}
\end{eqnarray}
The expression (\ref{eq:spectrum_general_movement}) is valid for an arbitrary motion
of the mirror, as long as it is sufficiently slow ($|\delta\dot q|\ll 1$) and has a
small amplitude. In the following section, we present our results for a
specific but very useful law of motion, that of a mirror which oscillates, practically, at a single
frequency.

\section{Results and Discussion}
\label{Results-and-Discussion}

Let us consider that the movement of the mirror is described by
\begin{equation}
\delta q(t)=\epsilon_{0}\cos(\omega_{0}t)e^{-|t|/\tau},
\label{eq:motion_of_mirror}
\end{equation}
which is a typical motion considered in investigations of the DCE \cite{MaiaNeto-Machado-1996},
where $\epsilon_{0}$ represents its amplitude, $\omega_0$ the dominating mechanical frequency and $\tau$
the effective time interval of the oscillation. The Fourier transform of Eq. (\ref{eq:motion_of_mirror}) is
\begin{equation}
\delta Q(\omega)=
\epsilon_0\tau\left[\frac{1}{1+(\omega+\omega_0)^2\tau^2}+\frac{1}{1+(\omega-\omega_0)^2\tau^2}\right].
\label{eq:deltaQ_Fourier}
\end{equation}
In the limit with $\omega_{0}\tau\gg1$, we obtain
\begin{equation}
|\delta Q(\omega)|^2 \approx \frac{\pi}{2}\epsilon_{0}^2\tau[\delta(\omega-\omega_0)+\delta(\omega+\omega_0)],
\label{eq:deltaQ_Fourier_undamped}
\end{equation}
which is reasonable once (\ref{eq:deltaQ_Fourier}) possesses two narrow peaks around
$\omega=\pm \omega_0$ in this limit.
Substituting (\ref{eq:deltaQ_Fourier_undamped}) into (\ref{eq:spectrum_general_movement}) and
defining $\tilde{{\cal N}}$ the normalized particle spectrum per unit area, per unit time,
per unit solid angle, given by
\begin{equation}
\tilde{{\cal N}}=
\frac{1}{A}\frac{dN}{d\omega d\Omega}
\left(\frac{\epsilon_0^2\tau}{8\pi^3}\right)^{-1},
\label{eq:}
\end{equation}
we can write
\begin{equation}
\tilde{{\cal N}}(\omega,\omega_0,\gamma,\theta)=\omega_0^4\;\cal{\tilde{F}}(\alpha,\beta,\theta)
\label{eq:spectrum_underdamped_cosine}
\end{equation}
where
\begin{eqnarray}
\cal{\tilde{F}}(\alpha,\beta,\theta) &=&
\frac{\alpha^3\cos^2\theta}{1+\alpha^2\beta^2\cos^2\theta}
\sqrt{(\alpha-1)^2-\alpha^2\sin^2\theta}
\frac{[1-\alpha^2\beta^2\sin^2\theta -\alpha\beta^2(1-\alpha)]^2}
{1+\beta^2[(1-\alpha)^2-\alpha^2\sin^2\theta]}
\times \cr &&
\Theta[1-\alpha(1+\sin\theta)],
\end{eqnarray}
with $\alpha=\omega/\omega_0$ and $\beta=\omega_0\gamma$.
Notice that the Heaviside function present in Eq. (\ref{eq:spectrum_underdamped_cosine})
indicates the following interesting features of the particle spectrum:
particles associated to a certain frequency $\omega$ can be created in all angles if
$0 \leq \omega \leq \omega_0/2$;
for particles associated to a frequency $\omega>\omega_0/2$, there is
no particle emission for angles larger than $\theta_0(\omega)$, with
\begin{equation}\label{eq:theta_0}
 \theta_0(\omega)=\arcsin\left(\frac{\omega_0-\omega}{\omega}\right).
\end{equation}
Moreover, observe that expression (\ref{eq:spectrum_underdamped_cosine}) coincides with the
results found in \cite{MaiaNeto-Machado-1996} in two particular cases. In
the limit $\gamma\rightarrow0$, Eq. (\ref{eq:spectrum_underdamped_cosine})
reproduces the particle spectrum for TE photons, while for
$\gamma\rightarrow\infty$ it corresponds to the spectrum of TM photons.

The normalized particle spectrum per unit area, per unit time,
here labeled as ${\cal{N}}$, can be
obtained by integrating (\ref{eq:spectrum_underdamped_cosine}) in the solid angle,
with angular range
$\theta\in[0,\pi/2]$:
\begin{equation}
{\cal{N}}(\omega,\omega_0,\gamma)=\omega_0^4\;{\cal{{F}}}(\alpha,\beta)
\label{sol}
\end{equation}
with
\begin{equation}
\;{\cal{F}}(\alpha,\beta)=2\pi\int_0^{\pi/2}\; d\theta\;\sin\theta\;\cal{\tilde{F}}(\alpha,\beta,\theta).
\label{lua}
\end{equation}
The results for $\omega_0=1$ and different values of $\gamma$ in ${\cal{N}}(\omega,\omega_0,\gamma)$ are shown in
Fig. 1. The spectrum is symmetric around $\omega_0/2$ for every value of $\gamma$.
This is in agreement with the expectation that the particles are created in pairs
with frequencies $\omega_1$ and $\omega_2$ so that $\omega_1+\omega_2=\omega_0$, the
driving frequency of the mirror \cite{MaiaNeto-Machado-1996}.

%
\begin{figure}[!hbt]
\begin{center}
\includegraphics[width=9cm,angle=0]{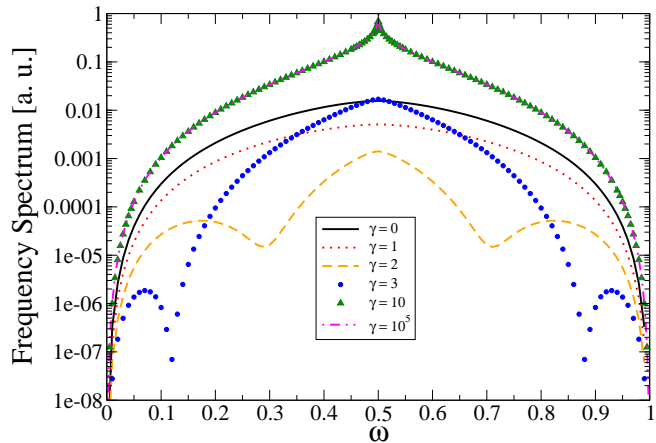}
\caption{(Color online.) The normalized spectrum ${\cal N}$
in arbitrary units, as function of $\omega$, for different values
of $\gamma$, with $\omega_0=1$.
The particle emission at a given
frequency $\omega$ is initially reduced as $\gamma$ increases
from zero until a given angle and then it starts increasing, reaching a fixed
plateau for $\gamma\rightarrow\infty$.
}
\end{center}
\label{fig-spectral-density}
\end{figure}
%

One can also analyze the angular spectrum of emitted particles. Our results for
$\cal{\tilde{N}}$, which follow from
Eq. (\ref{eq:spectrum_underdamped_cosine}) for a particular frequency $\omega=\omega_0/2$ and
$\omega_0=1$, are shown in Fig. 2. A few comments are in order. Firstly,
for small values of the Robin parameter, $\gamma\lesssim1$, the emission is mainly
forward, with large angle emission being suppressed. In second place, for $\gamma\simeq2$,
the emission is strongly inhibited for every angle, but it also acquires a maximum around $\theta\simeq1$rad.
If the $\gamma$ is further increased, the emission rate rises again as a whole,
but the particle production is highest around $\theta\simeq1$rad, and not around the normal direction
$\theta=0$.
Another interesting and subtle point becomes evident when we compare the angular spectrum in the
case of Robin boundary conditions with $\gamma\gg1$ and that of the Neumann BCs.
We see from Fig. 2 that there is no emission for $\theta\rightarrow\pi/2$ no matter how large
the values of $\gamma$ are. This is basically due to the factor $\cos^2\theta$ in
(\ref{eq:spectrum_underdamped_cosine}). However, as shown in \cite{MaiaNeto-Machado-1996},
there is a {\it finite} emission rate at $\theta\rightarrow\pi/2$ for Neumann BC. The apparent paradox
can be solved by noticing that the limits $\theta\rightarrow\pi/2$ and $\gamma\rightarrow\infty$ in Eq.
(\ref{eq:spectrum_general_movement}) do not commute.
Fortunately, this subtlety affects only grazing angles and the limit $\gamma\rightarrow\infty$
can be identified with the Neumann BC whenever $\theta=\pi/2$ is not the only angle in consideration.
%
\begin{figure}[!hbt]
\begin{center}
\includegraphics[width=9cm,angle=0]{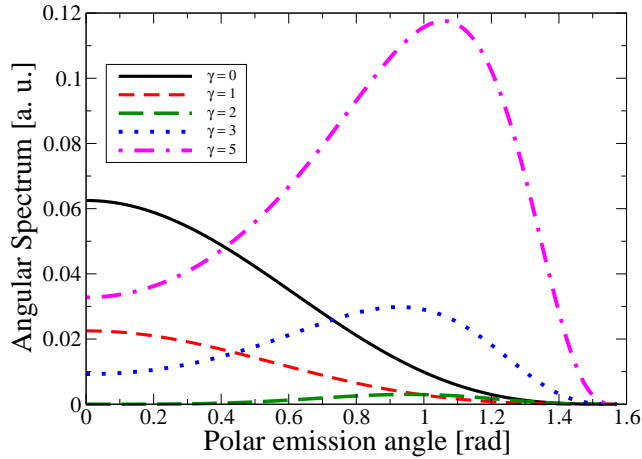}
\caption{(Color online.) Angular spectrum ${\cal{\tilde{N}}}$ of emitted particles with frequency $\omega=\omega_0/2$,
with $\omega_0=1$, as a function of the emission angle (measured with respect to the normal to the mirror)
for various values of $\gamma$. Solid (black) line: $\gamma=0$; dashed (red)
line: $\gamma=1$; long dashed (green) line : $\gamma = 2$;
dotted (blue) line: $\gamma=3$; dashed-dotted (magenta) line:
$\gamma=5$.
}
\end{center}
\label{fig1:polar_density}
\end{figure}

Finally, let us turn our attention to the total number of particles emitted $N$,
which is given by
\begin{equation}
N(\omega_0,\gamma)=\omega_0^5\;F(\beta)
\label{total-number}
\end{equation}
where
\begin{equation}
F(\beta)= \int_0^{\infty}\;d\alpha\;{\cal F}(\alpha,\beta).
\label{F-total-number}
\end{equation}
The ratio between the emission rate $N_R=N(\omega_0,\gamma)$
and the emission rate $N_D=N(\omega_0,0)$ (Dirichlet BC),
is given by
\begin{equation}
N_R/N_D=F(\beta)/F(0),
\label{ratio}
\end{equation}
which is showed in Fig. 3. As first demonstrated
in \cite{MaiaNeto-Machado-1996}, the total emission rate for Neumann BC is 11 times
larger than that for Dirichlet BC in $3+1$ dimensions.
Our results not only confirm the factor 11 but
also show that Dirichlet and Neumann cases are connected by a non-monotonic curve. Indeed, the Robin BC with
$\gamma\omega_0\simeq2$ provides a very strong inhibition of the particle
production. This property of the DCE with Robin BC had already been noticed in $1+1$ dimensions
\cite{Mintz-Farina-Maia-Neto-Robson-JPA-2006-II},
but it was not obvious {\it a priori} that this property would be preserved at higher dimensions.

%
\begin{figure}[!hbt]
\begin{center}
\includegraphics[width=9cm,angle=0]{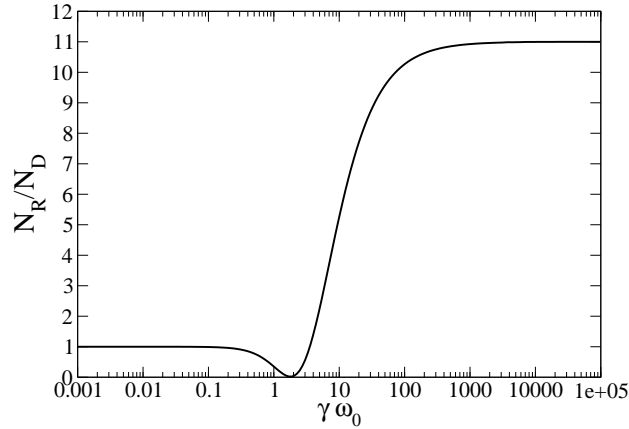}
\caption{(Color online.) Total emission rate for Robin BC
normalized by the Dirichlet rate. Notice that the Neumann limit
($\gamma\rightarrow\infty$) is correctly recovered. The strong decoupling
between the mirror and the quantum vacuum field at $\gamma\omega_0\simeq2$
is typical of the Robin BC \cite{Mintz-Farina-Maia-Neto-Robson-JPA-2006-I,Mintz-Farina-Maia-Neto-Robson-JPA-2006-II}.
}
\end{center}
\label{fig2:total_rate}
\end{figure}

\section{Final remarks}
\label{final-coments}

We have investigated the DCE for a real massless scalar field in $3+1$ dimensions satisfying a Robin
BC at a non-relativistic moving plate (in the tangential frame). We used the perturbative method of Ford and Vilenkin \cite{Ford-Vilenkin-PRD-1982}, valid for non-relativistic motions of small amplitudes to evaluated the Bogoliubov
transformations between creation and annihilation operators in the remote past, $a_{in}$, and
distant future, $a_{out}$. This allowed us to determine the spectral and angular distributions of the
created particles caused by the moving mirror. Eq. (\ref{eq:spectrum_general_movement})
 exhibits our expression for the particle spectrum per unit area,
valid for a general (non-relativistic) law of motion for the moving plate. 
Assuming the oscillating law of motion given by Eq. (\ref{eq:motion_of_mirror}) - a typical motion considered
in investigations of the DCE - we obtained an explicit expression for the spectrum per unit area, given by 
Eq. (\ref{eq:spectrum_underdamped_cosine}). In the limit $\gamma\rightarrow0$ (Dirichlet case) we recovered the particle spectrum for TE photons, whereas for $\gamma\rightarrow\infty$ (Neumann case) we reobtained the spectrum of emitted TM photons.

\noindent
Our results also show that, although in the limits $\gamma\rightarrow0$ and $\gamma\rightarrow\infty$ the total number of created particles is a monotonic function of the mechanical frequency of the plate,
$\omega_0$, the same is not true for intermediate values of $\gamma$. In fact, for any fixed positive $\gamma$, the total number of created particles is not a monotonic function of $\omega_0$. More than that, the strong inhibition of the DCE that occurs in $1+1$ dimensions, when the Robin BC is present, also occurs in $3+1$ dimensions. Indeed, the total number of created particles shown in Fig. 3 is dramatically reduced for a quasi-harmonic motion with a frequency $\omega_0$ such that $\gamma\omega_0\simeq2$, where $\gamma$ is the parameter that characterizes the Robin condition. 
Naively thinking, this surprising effective decoupling between the plate and the quantized field, predicted previously in $1+1$ dimensions \cite{Mintz-Farina-Maia-Neto-Robson-JPA-2006-I,Mintz-Farina-Maia-Neto-Robson-JPA-2006-II} was not expected in $3+1$, since in the latter case only the field modes that propagates perpendicularly to the plate are expected to behave as the field modes in $1+1$ dimensions. 
 
The suppression just discussed show how the dynamical Casimir effect may be strongly dependent on the boundary conditions employed. This fact is very important, since any extra information about the created particles, either in the total number of them or in the angular dependence of the corresponding spectral distribution, can be extremely useful to better identify the dynamical Casimir photons in a given experiment. Since the Robin parameter $\gamma$ can be interpreted as the plasma wavelength of a given material, the results presented here may be of some help for future experiments, at least as a source of concern and caution when dimensioning these experiments. For the moment, realistic values for the product $\gamma\omega_0$ are still very far from a strong suppression, but this may not be the case in future experiments.  It would be interesting if the peculiar signature of the DCE with Robin conditions -  the strong inhibition in the particle emission -  could eventually be captured in experiments.
 Considering the connections between Robin BC and the theoretical model
underlying the first experimental observation of the DCE \cite{C-Wilson-et-al-Nature-2011},
we believe that a thorough study of the implications of the Robin
boundary conditions on the DCE is crucial. 

As a final comment, since the dissipative force acting on a moving  mirror is closely related to the particle creation, we expect that this dissipative force will also suffer a similar inhibition in $3+1$ dimensions, as it occurs in the $1+1$ dimensional case \cite{Mintz-Farina-Maia-Neto-Robson-JPA-2006-I}. This calculation will be left for a future work.

\section*{Acknowledgments}

The authors thank H.O. Silva for valuable discussions and the Brazilian agencies CNPq, CAPES and FAPERJ for partial financial support. B.W.M. also thanks Universidade do Estado do Rio de Janeiro for support through the ``Professor Visitante'' fellowship. A.L.C.R. also
thanks Universidade Federal do Par\'a for the hospitality.



\begin{thebibliography}{99}

\bibitem{Moore-1970} G.T. Moore, J Math. Phys. {\bf 11}, 2679 (1970).

\bibitem{Dewitt-PhysRep-1975} B.S. DeWitt, Phys. Rep. {\bf 19}, 295 (1975).

\bibitem{Fulling-Davies-PRS-1976-I} S.A. Fulling and P.C.W. Davies, Proc. R. Soc. London, {\bf A 348}, 393 (1976).

\bibitem{Davies-Fulling-PRS-1977-I} P.C.W. Davies and S.A. Fulling, Proc. R. Soc. London, {\bf A 354}, 59 (1977).

\bibitem{Davies-Fulling-PRS-1977-II} P. C. W. Davies and S. A. Fulling, Proc. R. Soc. London, {\bf A 356}, 237 (1977).

\bibitem{Nyquist-1928}  H. Nyquist, {Phys. Rev.} {\bf 32}, 110-113 (1928).

\bibitem{Callen-Welton-1951}  H.B. Callen and T.A. Welton, {Phys. Rev.} {\bf 83}, 34 (1951).

\bibitem{Barton91} G. Barton, {J. Phys. A: Math. Gen.} {\bf 24}, 991 (1991).

\bibitem{Braginsky-Khalili-1991}  V.B. Braginsky and F.Ya. Khalili, {Phys. Lett.} {\bf 161}, 197 (1991).

\bibitem{Jaekel-Reynaud-1992}  M.T. Jaekel and S. Reynaud, {Quant. Opt.} {\bf 4}, 39 (1992).

\bibitem{Barton94} G. Barton, in Cavity Quantum Electrodyamics, Supplement: Advances in Atomic, Molecular and Optical Physics, edited by P. Berman, (Academic Press, New York, 1993).

\bibitem{MaiaNeto-Reynaud-1993}  P.A. Maia Neto and S. Reynaud, {Phys. Rev.} A {\bf 47}, 1639 (1993).

\bibitem{Mintz-Farina-Maia-Neto-Robson-JPA-2006-I} B. Mintz, C. Farina, P.A. Maia Neto and R.B. Rodrigues, {J. Phys.} A {\bf 39}, 6559 (2006).

\bibitem{Farina-BJP-2006} C. Farina, Braz. J. Phys. {\bf 36}, 1137 (2006).

\bibitem{MaiaNeto-1994} P.A. Maia Neto  J. Phys. A {\bf 27}, 2167 (1994).

\bibitem{MaiaNeto-Machado-1996} P.A. Maia Neto and L.A.S. Machado, {Phys. Rev.} A \textbf{54}, 3420 (1996).

\bibitem{MaiaNeto-Mundurain-1998} D.F. Mundarain and P.A. Maia Neto, Phys. Rev. A {\bf 57}, 1379 (1998).

\bibitem{Dodonov-Revisao} V.V. Dodonov, {Adv. Chem. Phys.} {\bf 192}, 309 (2001).

\bibitem{PAMN-EtAl-Revisao} D.A.R. Dalvit, P.A. Maia Neto and F.D. Mazzitelli, in {\it Casimir Physics}, edited by D.A.R. Dalvit, P. Milonni, D. Roberts and F. da Rosa, Lecture Notes in Physics, Vol. 834 (Springer, New York, 2011).

\bibitem{V-V-Dodonov-Phys-Scrip-2010}  V.V. Dodonov, {\it Phys. Scrip.} {\bf 82}, 038105 (2010).

\bibitem{Braggio-Agnesi} C. Braggio {\it et al}, {Europhys. Lett} {\bf 70}, 754 (2005);
A. Agnesi {\it et al}, {J. Phys.} A {\bf 41}, 164024 (2008);
A. Agnesi {\it et al}, {J. Phys: Conf. Series} {\bf 161}, 012028 (2009).

\bibitem{Yablonovitch-1989} E. Yablonovitch, {Phys. Rev. Lett.} {\bf 62}, 1742 (1989).

\bibitem{LozovikEtAl-1995} Yu.E. Losovik, V.G. Tsvetus and E.A. Vinogradov, {JETP Lett.} {\bf 61} 723 (1995).

\bibitem{Kim-Brownell-Onofrio-EPL-2007} W.-J. Kim, J.H. Brownell and R. Onofrio , {Europhys. Lett} {\bf 78}, 21002 (2007).

\bibitem{Dodonov-2005} V.V. Dodonov, {J. Opt.} B {\bf 7}, S445 (2005).

\bibitem{Dodonov-2009} V.V. Dodonov, {Phys. Rev.} A {\bf 80}, 023814-1 (2009).

\bibitem{Kim-Brownell-Onofrio-PRL-2006} W.J. Kim, J.H. Brownell and R. Onofrio , {Phys. Rev. Lett.} {\bf 96}, 200402 (2006).

\bibitem{Dezael-Lambrecht-EPL-2010} F.X. Dezael and A. Lambrecht, {Eur. Phys. Lett.} {\bf 89}, 14001 (2010).

\bibitem{Kawakubo-Yamamoto-PRA-2011} T. Kawakubo and K. Yamamoto, {Phys. Rev.} A, {\bf 83}, 013819 (2011).

\bibitem{Faccio-Carusotto-EPL-2011}  D. Faccio and I. Carusotto, {Eur. Phys. Lett} {\bf 96}, 24006 (2011).

\bibitem{C-Wilson-et-al-Nature-2011} C. M. Wilson, G. Johansson, A. Pourkabirian, M. Simoen, J. R. Johansson, T. Duty, F. Nori and P. Delsing,
 {Nature}, {\bf 479}, 376 (2011).

\bibitem{J-R-Johansson-G-Johansson-C-Wilson-F-Nori-PRL-2009} J.R. Jo\-han\-sson, G. Jo\-han\-sson, C.M. Wilson and F. Nori,
 {Phys. Rev. Lett.}  {\bf 103}, 147003 (2009).

\bibitem{Silva-Farina-PRD-2011} Hector O. Silva and C. Farina, {Phys. Rev.} D, {\bf 84}, 045003 (2011).

\bibitem{Farina-Silva-Rego-Alves-IJMPCS-2012} C. Farina, Hector O. Silva, Andreson L. C. Rego and Danilo T. Alves,
 {Int. J. Mod. Phys. Conf. Ser.}, {\bf 14}, 306 (2012).

\bibitem{Mintz-Farina-Maia-Neto-Robson-JPA-2006-II} B. Mintz, C. Farina, P.A. Maia Neto and R.B. Rodrigues, { J. Phys.} A {\bf 39}, 11325 (2006).

\bibitem{Lambrecht-PRL-1996} A. Lambrecht, M.-T. Jaekel and S. Reynaud, Phys. Rev. Lett. {\bf 77}, 615 (1996).

\bibitem{Diogo-Farina-Procceeding-2011} Diogo Azevedo, F. Pascoal and C. Farina in: {\it Proceedings of the IX Workshop on Quantum Field Theory under the Influence of External Conditions}, edited by K.A. Milton and M. Bordag, (World Scientific, Singapore, 2009).

\bibitem{Bogoliubov} N. N. Bogoliubov, {Sov. Phys.} JETP, {\bf 7}, 51 (1958).

\bibitem{Ford-Vilenkin-PRD-1982} L.H. Ford and A. Vilenkin, {Phys. Rev.} D {\bf 25}, 2569 (1982).


\end{thebibliography}
\end{document}